\RequirePackage[hyphens]{url}
\documentclass[journal=jctc,manuscript=article]{achemso}
\setkeys{acs}{articletitle=true}

\pdfoutput=1

\usepackage{hyperref,url}
\usepackage{amsmath}
\usepackage{amsfonts}
\usepackage[capitalise]{cleveref}
\usepackage{placeins}
\hypersetup{breaklinks,colorlinks=true,linkcolor=blue,citecolor=blue,urlcolor=blue,filecolor=blue}
\usepackage{graphicx}
\usepackage{dcolumn}

\newcommand{\br}{\mathbf{r}}
\newcommand{\bx}{\mathbf{x}}

\newcommand{\bG}{\mathbf{G}}
\newcommand{\hh}{\hat{H}}

\newcommand{\hc}{\hat{c}}
\newcommand{\hcd}{\hat{c}^{\dagger}}

\author{Fionn D. Malone}
\affiliation{Quantum Simulations Group, Lawrence Livermore National Laboratory, 7000 East Avenue, Livermore, CA, 94551 USA.}
\email{malone14@llnl.gov}
\author{Shuai Zhang}
\affiliation{Quantum Simulations Group, Lawrence Livermore National Laboratory, 7000 East Avenue, Livermore, CA, 94551 USA.}
\author{Miguel A. Morales}
\affiliation{Quantum Simulations Group, Lawrence Livermore National Laboratory, 7000 East Avenue, Livermore, CA, 94551 USA.}
\email{moralessilva2@llnl.gov}

\title{Overcoming the Memory Bottleneck in Auxiliary Field Quantum Monte Carlo Simulations with Interpolative Separable Density Fitting}

\begin{document}
\begin{abstract}
We investigate the use of interpolative separable density fitting (ISDF) as a means to reduce the memory bottleneck in auxiliary field quantum Monte Carlo (AFQMC) simulations of real materials in Gaussian basis sets.
We find that ISDF can reduce the memory scaling of AFQMC simulations from $\mathcal{O}(M^4)$ to $\mathcal{O}(M^2)$. 
We test these developments by computing the structural properties of Carbon in the diamond phase, comparing to results from existing computational methods and experiment.
\mbox{(LLNL-JRNL-759024)}
\end{abstract}

\date{\today}
\maketitle

\section{Introduction}
The accurate, parameter free description of materials has been a grand challenge of electronic structure theory for decades.
Density functional theory (DFT) is by far the most widely used first-principles based approach to materials science, combining an often good enough accuracy with a modest computational cost.
However, DFT results can depend sensitively on the choice of approximate exchange correlation functional.
Calculating band gaps in semiconductors\citep{QPSCGW}, discerning different phases in hydrogen under extreme conditions\cite{MoralesNQE,Azadi2013,Clay2014}, and the systematic description of strongly correlated materials\citep{DMCNio} are some well known limitations of DFT.
The development of hybrid functionals\cite{HSE2006,pbe0a,pbe0b,b3lyp1,b3lyp2} and adaptations of the exchange-correlation functional for strong correlations\cite{LDAU1998} can sometimes improve results, but require an often ad-hoc determination of additional parameters.

Wavefunction based quantum chemical methods offer an alternative, systematic approach to solving the electronic structure problem directly. 
Unfortunately, they come with a cost which is often prohibitively large.
For example, standard Coupled-Cluster theory (including single and double excitations) scales like the sixth power of the system size, while the exact approach of FCI scales exponentially.
Nevertheless, they are increasingly and successfully being applied to problems in solid state physics\citep{Stoll09,gillan_high-precision_2008,Manby09,booth_towards_2013,mcclain_gaussian-based_2017,SunDF17}.

Quantum Monte Carlo (QMC) methods offer another route to directly solving the many-electron Schr{\"o}dinger equation, with often much more favorable scaling.
Real space diffusion Monte Carlo \citep{RevModPhys.73.33} is perhaps the most widely used QMC approach to solving electronic structure problems, and can now routinely simulate 100s to 1000s of atoms, making full use of modern supercomputers\cite{qmcpack}.
In order to overcome the fermionic sign problem, DMC uses a trial wavefunction to impose the fixed-node approximation.
Although results for the uniform electron gas suggest that the fixed-node approximation is extremely accurate\citep{PhysRevB.85.081103}, it is often difficult to improve the nodes for more realistic systems, where the nodal error can be more significant.
Additionally, non-local pseudopotentials, which are eventually necessary for describing heavier elements, are difficult to use and require additional uncontrolled approximations (e.g. the locality approximations\cite{MitasLocality}) which are also hard to assess and improve.

Phaseless auxiliary field QMC \citep{PhysRevLett.90.136401,mottareview} (AFQMC) offers an alternative route to overcoming some of these issues. Similar to DMC, it also uses a constraint to remove the fermion sign problem at the expense of a bias\citep{PhysRevLett.90.136401}, which it implements with a trial wavefunction. Multi-determinant\citep{PhysRevB.94.085103,borda2018non}, generalized Hartree--Fock \citep{PhysRevB.94.085103,chang2017multi}, and self-consistently determined trial wavefunctions\citep{PhysRevB.94.235119,Zheng1155} have been found to dramatically improve the phaseless error while only modestly increasing the computational effort.
AFQMC also works in a second-quantized orbital-based basis, allowing for the easier evaluation of ground state properties other than the total energy.
For example, dipole moments and reduced density matrices\citep{doi:10.1021/acs.jctc.7b00730}, excited states\citep{1367-2630-15-9-093017,doi:10.1063/1.4861227,PhysRevB.94.085140} and forces \citep{mottaforces} can now be routinely calculated.
Additionally, non-local pseudpotentials\citep{PhysRevB.75.245123,1367-2630-15-9-093017} and spin-orbit coupling operators\citep{RosenberSpinOrbitAFQMC} can be straightforwardly incorporated and require no additional approximations.

One of the main limiting factors in the use of AFQMC for real materials is its large storage requirements, driven by the need of the two-electron integral tensor, which also plagues many quantum chemistry methods.
Naively, this leads to an $\mathcal{O}(M^4)$ storage requirement. Even with the use of factorizations based on Cholesky decomposition\citep{mchol0,mchol,mchol2,doi:10.1063/1.3654002} or density fitting\citep{WhittenDF73}, which can reduce the storage requirements to $\mathcal{O}(M^3)$ (at the expense of a much higher computational cost when evaluating the total energy), the resulting approach is still too expensive for large systems and so far is typically limited to systems with 30-40 atoms. 

In this paper we discuss a way to reduce the memory requirement of AFQMC to $\mathcal{O}(M^2)$ using tensor factorization approaches, in particular using interpolative separable density fitting (ISDF)\citep{LU2015329,doi:10.1021/acs.jctc.7b00807,doi:10.1021/acs.jctc.7b01113}. We first review the basics of AFQMC and its standard implementation. Next we describe the new formulation of AFQMC based on ISDF and its implementation in the open-source QMCPACK software package \citep{qmcpack}.
Finally, we benchmark these developments by computing the equation of state and cohesive energy of Carbon in the diamond phase near ambient conditions for simulation cells containing up to 128 atoms in the TZVP basis set.

\section{Methods}
In this section we will outline the basics of the AFQMC method in order to introduce the relevant notation necessary to describe the ISDF procedure.
We will focus on simulating real solids in quantum chemical (periodic Gaussian) basis sets.

\subsection{Introduction to AFQMC\label{sec:intro_afqmc}}
The many-electron Hamiltonian describing a collection of electrons and static nuclei can be written in second quantized form as:

\begin{align}
    \hat{H} &= \sum_{ij\sigma} h_{ij} \hcd_{i\sigma}\hc_{i\sigma} + \frac{1}{2}\sum_{ijkl\sigma\sigma'}v_{ijkl}  \hcd_{i\sigma}\hcd_{j\sigma'}\hc_{l\sigma'}\hc_{k\sigma}+E_{II},\label{eq:hamil}\\
            &= \hh_1 + \hh_2 + E_{II},
\end{align}
where $E_{II}$ is the ionic energetic contribution, $\hcd_{i\sigma}$ and $\hc_{i\sigma}$ create and annihilate an electron in some single-particle state $|i\sigma\rangle$ and $\sigma$ is the electron's spin. The matrix elements of the one- and two-body parts of the Hamiltonian are given (in Hartree atomic units) as
\begin{equation}
h_{ij} = \int d\br \  \varphi_{i}^*(\br)\left(-\frac{1}{2}\hat{\nabla}_\br^2 + V_{e-I}(\br)\right)\varphi_{j}(\br),
\end{equation}
where $\langle \br | i\rangle = \varphi_{i}(\br)$ is the real-space representation of the ith single-particle state and $V_{e-I}(\br)$ is the electron-ion interaction. The electron-repulsion integrals (ERIs) are given by:
\begin{equation}
    v_{ijkl} = \int \int d\br \ d\br' \ \varphi^*_i(\br)\varphi^*_j(\br')\frac{1}{|\br-\br'|}\varphi_{k}(\br)\varphi_l(\br')\label{eq:four_ix}.
\end{equation}

Like most ground state QMC methods, AFQMC is a projector method relying on the fact that
\begin{equation}
|\Psi_0\rangle \propto \lim_{\tau\rightarrow\infty} e^{-\tau\hat{H}}|\phi\rangle,\label{eq:projection}
\end{equation}
where $|\Psi_0\rangle$ is the ground state of $\hh$ and $|\phi\rangle$ is some initial state satisfying $\langle \phi | \Psi_0\rangle \ne 0$.
The long time limit of \cref{eq:projection} is in practice found iteratively using
\begin{equation}
    |\Psi^{(n+1)}\rangle = e^{-\Delta\tau\hh} |\Psi^{(n)}\rangle\label{eq:iterate},
\end{equation}
where $\Delta\tau$ is the time step.
To evaluate the matrix exponentials in \cref{eq:iterate} we take $\Delta\tau$ to be small and use the symmetrized Suzuki-Trotter decomposition
\begin{equation}
e^{-\Delta\tau\hat{H}} = e^{-\frac{\Delta\tau}{2} \hat{H}_1}e^{-\Delta\tau \hat{H}_2}e^{-\frac{\Delta\tau}{2} \hat{H}_1} + \mathcal{O}(\Delta\tau^2)\label{eq:suz_trot}.
\end{equation}
In AFQMC we choose the many-particle states to be Slater determinants.
From Thouless' theorem\citep{thouless_theorem,thouless_theorem_2} we know that the exponential of a one-body operator applied to a Slater determinant yields another single Slater Determinant.
However, in order to realize the iterative solution given in \cref{eq:iterate} we need to apply the exponential of a two-body operator, which is difficult to do in general.
To make headway, we first write $\hh_2$ as a sum of squares of one-body Hamiltonians (specific formulae given below)
\begin{equation}
    \hat{H}_2 = \hat{v}_{0} + \frac{1}{2} \sum_{\gamma} \hat{v}_{\gamma}^2 \label{eq:hsq},
\end{equation}
and use the Hubbard-Stratonovich transformation
\begin{equation}
e^{-\frac{\Delta\tau}{2}\sum_{\gamma}\hat{v}_{\gamma}^2} = \prod_\gamma \int d x_\gamma e^{-\frac{x_\gamma^2}{2}} e^{\sqrt{-\Delta\tau} x_\gamma\hat{v}_\gamma} \label{eq:hs}.
\end{equation}
Inserting \cref{eq:hs} into \cref{eq:suz_trot}, we can rewrite \cref{eq:iterate} as
\begin{equation}
    |\Psi^{(n+1)}\rangle =   \int d \bx p(\bx)\hat{B}(\bx) |\Psi^{(n)}\rangle, \label{eq:iterate_hs}
\end{equation}
where $\hat{B}(\bx)$ now contains exponentials of one-body operators only.
The multi-dimensional integral in \cref{eq:iterate_hs} can be evaluated using Monte Carlo integration over normally distributed auxiliary fields $\bx$.
However, with an eye to eventually imposing a constraint\citep{PhysRevB.55.7464}, the projection to the ground state is instead performed using an open-ended random walk wherein a set of weighted random walkers ($|\Psi^{(n)}\rangle = \sum_\alpha w^{(n)}_\alpha|\phi^{(n)}_\alpha\rangle$) representing the Slater determinants evolve according to \cref{eq:iterate_hs}.
Although this `free projection' algorithm is exact it suffers from a serious phase problem.
As the projection proceeds the walkers will acquire a phase which in the long imaginary time limit will be uniformly distributed in the complex plane rendering the accumulation of statistics impossible.
In order to overcome this issue Zhang \emph{et al.} introduced importance sampling and the `phaseless' approximation\citep{PhysRevLett.90.136401} which, at the cost of introducing a systematic bias, removes the instabilities associated with the fermion sign problem.

When using importance sampling, the walker's are propagated according to the modified propagator
\begin{equation}
    w_\alpha^{(n+1)}|\phi_\alpha^{(n+1)}\rangle =   \left[\int d \bx p(\bx) I(\bx,\bar{\bx},|\phi\rangle) \hat{B}(\bx-\bar{\bx})\right]w_\alpha^{(n)}|\phi^{(n)}\rangle, \label{eq:iterate_imp}
\end{equation}
where
\begin{equation}
I(\bx,\bar{\bx},|\phi\rangle) = \frac{\langle \psi_T|\hat{B}(\bx- \bar{\bx}) |\phi\rangle}{\langle \psi_T|\phi\rangle} e^{\bx\cdot\bar{\bx}-\frac{\bar{\bx}\cdot\bar{\bx}}{2}}
\end{equation}
is the importance function, $\bar{\bx}$ is the `force-bias' shift and $|\psi_T\rangle$ is a trial wavefunction.
Although the importance function encourages walkers to areas of Hilbert space with a larger overlap with the trial wavefunction, its primary purpose is to impose a constraint.
To control the phase problem the walker's Slater determinants are propagated as in free projection, but now their weights are constrained via
\begin{equation}
    w_\alpha^{(n+1)} = |I(\bx,\bar{\bx},|\phi_\alpha^{(n)}\rangle)|\times \max \left(0, \cos \Delta \theta\right) w_\alpha^{(n)},
\end{equation}
where the phase is defined as
\begin{equation}
    \Delta \theta = \arg\left(\frac{\langle \psi_T|\hat{B}(\bx- \bar{\bx}) |\phi\rangle}{\langle \psi_T|\phi\rangle}\right).
\end{equation}
The trial wavefunction enforces the `phaseless' constraint, which produces exact results if $|\Psi_T\rangle=|\Psi_0\rangle$.
With this approximation, the walkers' weights remain real and positive and those walkers with rapidly changing phases are killed and removed from the simulation\citep{hybrid}.

\subsection{Practical Implementation}
A key step in the AFQMC algorithm is the factorization of the two-body Hamiltonian which is necessary in order to perform the Hubbard-Stratonovich transformation.
We begin by rewriting the many-electron Hamiltonian as

\begin{align}
    \begin{split}
\hat{H} = &\sum_{ij\sigma} \left(h_{ij}-\frac{1}{2}\sum_{k} v_{ikkj}\right) \hcd_{i\sigma}\hc_{j\sigma} \\
        &+ \frac{1}{2}\sum_{ikjl\sigma\sigma'}v_{ijkl}  \hcd_{i\sigma}\hc_{k\sigma}\hcd_{j\sigma'}\hc_{l\sigma'},\\
        = &\hh_1' + \hh_2',
    \end{split}
\end{align}
where we have omitted any constant factors.
We next need to factorize the ERIs in order to write $\hh_2'$ as a sum of squares of one-body operators.
Any exact representation of $v_{ijkl}$ requires storing an impractical $\mathcal{O}(M^4)$ complex numbers so a more compact representations is desired.
Previous AFQMC studies using Gaussian basis sets have predominantly relied on a modified Cholesky decomposition\citep{mchol0,mchol,mchol2,doi:10.1063/1.3654002} to write the ERIs as (in general)
\begin{equation}
    v_{ijkl} \approx \sum_\gamma^{N_\gamma} L_{ik}^{\gamma} L^{\gamma*}_{lj},
\end{equation}
where $N_\gamma = c_\gamma M$ is the number of Cholesky vectors necessary to reproduce the ERIs to within a given threshold.

With this factorization of the ERIs we can now introduce the Hermitian operators

\begin{align}
    \hat{v}_{\gamma+} &= \sum_{ik\sigma} \left(\frac{L^{\gamma}_{ik}+L^{\gamma*}_{ki}}{2}\right)\hcd_{i\sigma}\hc_{k\sigma} \\
    &=  \sum_{ik\sigma}\left[L_{+}\right]_{ik}^{\gamma}\hcd_{i\sigma}\hc_{k\sigma}, \\
\hat{v}_{\gamma-} &= i\sum_{ik\sigma} \left(\frac{L^{\gamma}_{ik}-L^{\gamma*}_{ki}}{2}\right)\hcd_{i\sigma}\hc_{k\sigma}\\
&=  \sum_{ik\sigma}\left[L_{-}\right]_{ik}^{\gamma}\hcd_{i\sigma}\hc_{k\sigma},
\end{align}
so that we can write
\begin{equation}
    \hh_2' = \frac{1}{2} \sum_\gamma \left(\hat{v}^{2}_{\gamma+} +\hat{v}^2_{\gamma-}\right),
\end{equation}
leading to $2c_\gamma M$ auxiliary fields.

Another crucial component for the practical implementation of the phaseless AFQMC method is the `force-bias' shift $\bar{\mathbf{x}}$.
One can show\citep{PhysRevLett.90.136401} that the optimal force-bias which cancels fluctuations in the importance function to $\mathcal{O}(\sqrt{\Delta\tau})$ can be written as
\begin{equation}
    \bar{x}_\gamma = -\sqrt{\Delta\tau}\frac{\langle \Psi_T| \hat{v}_\gamma | \phi \rangle}{\langle \Psi_T | \phi \rangle}.\label{eq:force_bias}
\end{equation}
This can now be evaluated in terms of the Cholesky vectors as
\begin{equation}
    \bar{x}^{\alpha}_{\gamma\pm} = -\sqrt{\Delta\tau}\sum_{ik\sigma}\left[L_{\pm}\right]^{\gamma}_{ik}G^{\alpha}_{i\sigma k\sigma},
\end{equation}
where the walker's Green's function has been defined as

\begin{align}
    G^{\alpha}_{i\sigma j\sigma'} &= \frac{\langle\psi_T|\hcd_{i\sigma}\hc_{j\sigma'}|\phi_\alpha\rangle}{\langle\psi_T|\phi_\alpha\rangle}\\
                           &= \left[U_{\sigma'}(V_\sigma ^{\dagger}U_{\sigma'})^{-1}V_{\sigma}^{\dagger}\right]_{ji}\\
    &=\left[V_\sigma^*(U_{\sigma'}^T V_\sigma^*)^{-1}U^{T}_{\sigma'}\right]_{ij}\label{eq:gf},
\end{align}
where $U_\sigma$ and $V_\sigma$ are the $M\times N_\sigma$ matrices of orbital coefficients for the walker $|\phi\rangle$ and trial wavefunction $|\psi_T\rangle$ respectively.
To reduce the cost of evaluating the force-bias shift we first precompute some tensors\citep{mottareview}.
If we write the Green's function in \cref{eq:gf} as

\begin{align}
    G_{i\sigma j\sigma'}^{\alpha} &= \left[V^{*}_\sigma \mathcal{G_{\sigma\sigma'}} \right]_{ij}
\end{align}
and define the partially contracted Cholesky vector
\begin{equation}
    \left[\mathcal{L}_{\pm}\right]^{\gamma}_{ak\sigma} = \sum_i \left[V^{*}_\sigma\right]_{ia}\left[L_{\pm}\right]^{\gamma}_{ik},
\end{equation}
then we can write\cite{mottareview}
\begin{equation}
    \bar{x}_{\gamma\pm} = -\sqrt{\Delta\tau}\sum_{ak\sigma}\left[\mathcal{L}_{\pm}\right]^{\gamma}_{ak\sigma}\mathcal{G}_{a\sigma k\sigma}.
\end{equation}
This brings the cost of computing the force-bias down from $\mathcal{O}(N_\gamma M^2)$ to $\mathcal{O}(N_\gamma NM)$ since $\mathcal{L}_{\pm}^{\gamma}$ can be computed once at the start of the simulation at the cost of $\mathcal{O}(N_\gamma M)$ operations.

Once the system has equilibrated we will have a statistical representation of the importance-sampled approximate ground state wavefunction
\begin{equation}
    |\Psi_0\rangle = \sum_\alpha w_\alpha \frac{|\phi_\alpha\rangle}{\langle \Psi_T|\phi_\alpha\rangle},
\end{equation}
from which we can compute estimates of observables.
The mixed estimator for the total energy is thus given as

\begin{align}
    E_{\mathrm{mixed}} &= \frac{\langle \psi_T | \hat{H} |\Psi_0\rangle}{\langle \Psi_T| \Psi_0\rangle} \\
                       &= \frac{\sum_\alpha w_\alpha E_L[\phi_\alpha]}{\sum_{\alpha}w_\alpha},
\end{align}
where
\begin{equation}
    \begin{split}
        E_L[\phi_\alpha] &= \sum_{ij\sigma} h_{ij}G^{\alpha}_{i\sigma j\sigma} + \\
        &\frac{1}{2}\sum_{ijkl\sigma\sigma'} v_{ijkl} \left(G^{\alpha}_{i\sigma
                k\sigma}G^{\alpha}_{j\sigma' l\sigma'}-G^{\alpha}_{i\sigma
                l\sigma'}G^{\alpha}_{j\sigma' k\sigma}\right)\label{eq:local_energy} \\ 
        &= E_{1B} + E_{2B},
    \end{split}
\end{equation}
is the walker's local energy.

To ensure the efficient evaluation of the two-body part of the energy we first
pre-contract the ERIs with the trial wavefunction to form
\begin{equation}
    \begin{split}
    \mathcal{V}_{(ak),(bl)}^{\sigma\sigma'} = \sum_{\gamma} \sum_{ij} L_{ik}^{\gamma} L_{lj}^{\gamma *} \Big(&\left[V^{*}_\sigma\right]_{ia} \left[V^{*}_{\sigma'}\right]_{jb}  - \\
    & \delta_{\sigma\sigma'}\left[V^{*}_\sigma\right]_{ib} \left[V^{*}_{\sigma'}\right]_{ja}\Big)\label{eq:tensor_v}.
    \end{split}
\end{equation}
The tensor $\mathcal{V}$ is computed once at the start of a simulation at the cost of $\mathcal{O}(N^2 M^3)$ 
operations and requires at most the storage of $2 N^2 M^2$ elements throughout the simulation.
The two-body contribution to the total energy can now be computed as
\begin{equation}
        E_{2B} = \frac{1}{2}\sum_{abkl\sigma\sigma'} \mathcal{V}^{\sigma\sigma'}_{(ak),(bl)}\mathcal{G}^\alpha_{a\sigma k\sigma}\mathcal{G}^{\alpha}_{b\sigma' l\sigma'}
\end{equation}
at the cost of $\mathcal{O}(N^2M^2)$ operations.
Back propagation can be used to compute expectation values of operators which do not commute with the Hamiltonian\cite{PhysRevB.55.7464,PhysRevE.70.056702,doi:10.1021/acs.jctc.7b00730}.

We can see from the discussion above that this straightforward implementation of the AFQMC algorithm could be prohibitively expensive for large problem sizes.
The fundamental limitation arises from the energy evaluation whose memory and computational cost scales quartically with the system size.
With no further optimization AFQMC would practically be limited to simulating relatively small simulation cells.
Fortunately, the Cholesky vectors $L_{ik}^{\gamma}$ and integral tensors are typically sparse.
This sparsity is enhanced due to fundamental symmetries of the Hamiltonian which can reduce the computational cost of AFQMC significantly.
For example, when simulating periodic solids, Bloch's theorem ensures that the number of non-zero matrix elements is reduced by a factor of $N_k$ where $N_k$ is the number of $k$-points.
Exploiting this fact ensures the cost (both memory and computational) of AFQMC simulations is cubic in the system size. 

\subsection{Interpolative Separable Density Fitting\label{sec:tensor}}

Although the modified Choleksy approach sketched out above has been successfully applied in AFQMC studies of molecules and solids, it is largely reliant on symmetries in order to ameliorate the memory overhead.
However, for systems with reduced symmetry, a larger unit cell size and for the computation of properties requiring a supercell approach (for example, formation energies of defects in solids), this memory overhead fundamentally limits the scope of AFQMC in Gaussian basis sets.
Thus, we seek a lower rank representation for the ERIs to overcome this issue.
In 2016, Lu and Ying introduced interpolative separable density fitting (ISDF) \citep{LU2015329}, which is similar in spirit to the tensor hypercontraction (THC) approach of Refs.\citenum{doi:10.1063/1.4732310,doi:10.1063/1.4768233,doi:10.1063/1.4768241}, that offers just this.
The ISDF approach has so far been applied to speeding up develop a cubic scaling RPA algorithm\citep{LU2017187}, speeding up hybrid functional calculations in DFT\citep{doi:10.1021/acs.jctc.7b00807,doi:10.1021/acs.jctc.7b01113} and the computation of excited state properties using the Bethe-Salpether approach\citep{hu2018accelerating}.
Here our aim is to adapt it to AFQMC.
We will focus on using the centroidal Veronoi tesselation ISDF developed in Ref.\citenum{doi:10.1021/acs.jctc.7b01113} with our notation and implementation closely following that found in Refs.~\citenum{doi:10.1021/acs.jctc.7b00807} and \citenum{doi:10.1021/acs.jctc.7b01113}.

The ISDF approach amounts to approximating the orbital products appearing in \cref{eq:hamil} as
\begin{equation}
    \varphi^{*}_i(\br)\varphi_k(\br) \approx \sum_\mu^{N_\mu} \zeta_\mu(\br) \varphi^*_i(\br_\mu)\varphi_k(\br_\mu),\label{eq:orb_prod}
\end{equation}
where the orbitals $\varphi_i(\br)$ are evaluated on a dense real space grid $\{\br_i\}_{i=1}^{N_g}$, $\{\br_\mu\}_{\mu=1}^{N_\mu}$ are a subset of these grid points, called the interpolating points, $\zeta_\mu(\br)$ are a set of interpolating vectors and $N_\mu = c_\mu M$ is the number of interpolating points.
Importantly, previous results suggest that $c_\mu$ is roughly independent of the system size, and a value in the range between $10-15$ is usually sufficient to reproduce total energies to sub mHa/atom accuracy\citep{doi:10.1021/acs.jctc.7b00807,doi:10.1021/acs.jctc.7b01113,hu2018accelerating}.
Inserting \cref{eq:orb_prod} into \cref{eq:four_ix} we find
\begin{equation}
    v_{ijkl} \approx \sum_{\mu\nu} \varphi^{*}_i(\br_\mu)\varphi_k(\br_\mu) M_{\mu\nu} (\varphi^{*}_l(\br_\nu)\varphi_j(\br_\nu))^{*}\label{eq:4ix_thc},
\end{equation}
where
\begin{equation}
    M_{\mu\nu} = \int d\br\br' \zeta_\mu(\br)K(\br,\br')\zeta^{*}_\nu(\br'),
\end{equation}
and $K(\br,\br')$ is the periodic Ewald potential.
Importantly (from the perspective of AFQMC), we see that the dominant storage requirement for representing the ERIs has been brought down to $\mathcal{O}(M^2)$.
The interpolating points can be found using the K-Means algorithm outlined in Ref.\citenum{doi:10.1021/acs.jctc.7b01113} and the interpolating vectors from the least squares solution to \cref{eq:orb_prod}.
Using the fact that the interpolating vectors are supercell periodic, we can then compute $M_{\mu\nu}$ as
\begin{equation}
    M_{\mu\nu} = \frac{4\pi}{\Omega} \sum_{\bG\ne\mathbf{0}} \frac{\zeta_\mu(\bG)\zeta^{*}_\nu(\bG)}{\bG^2},\label{eq:muv}
\end{equation}
where $\bG$ is a supercell reciprocal lattice vector and $\Omega$ is the supercell volume.
The interpolating vectors in reciprocal space can be efficiently computed using fast-fourier transforms\citep{frigo1998fftw}.

We next factorize\citep{factorize_M} $M= LL^{\dagger}$ and write
\begin{align}
\begin{split}
    \hh_2' =
    \frac{1}{2}\sum_\gamma \left(\sum_{\mu ik\sigma} \varphi_{i}^*(\br_\mu)\varphi_k(\br_\mu)L_{\mu\gamma}\hcd_{i\sigma}\hc_{k\sigma} \right)\times \\
        \left(\sum_{\nu jl\sigma'} \varphi_{j}^*(\br_\nu)\varphi_l(\br_\nu)L_{\gamma\nu}^*\hcd_{j\sigma'}\hc_{l\sigma
        '} \right),\\
    \end{split}\\
            &= \frac{1}{2}\sum_\gamma \hat{\rho}_\gamma \hat{\rho}^{\dagger}_\gamma\label{eq:thc_hamil_herm}.
\end{align}
To bring \cref{eq:thc_hamil_herm} into the form of a sum of squares of one-body operators we again define the Hermitian operators

\begin{align}
    \hat{v}_{\gamma+} &= \frac{1}{2} (\hat{\rho}_\gamma + \hat{\rho}^{\dagger}_\gamma) \\
    &=  \sum_{\mu ik\sigma} \varphi_{i}^*(\br_\mu)\varphi_k(\br_\mu)\mathrm{Re}\left[L_{\mu\gamma}\right]\hcd_{i\sigma}\hc_{k\sigma},\\
\hat{v}_{\gamma-} &=  -\frac{i}{2} (\hat{\rho}_\gamma - \hat{\rho}^{\dagger}_\gamma)\\
&= \sum_{\mu ik\sigma} \varphi_{i}^*(\br_\mu)\varphi_k(\br_\mu)\mathrm{Im}\left[L_{\mu\gamma}\right]\hcd_{i\sigma}\hc_{k\sigma},
\end{align}
where we have used

\begin{align}
    \varphi_{i}^*(\br_\mu)\varphi_k(\br_\mu)\left[L_{\mu\gamma} + L_{\mu\gamma}^*\right] &=
        2 \varphi_{i}^*(\br_\mu)\varphi_k(\br_\mu)\mathrm{Re}\left[L_{\mu\gamma}\right],\\
    \varphi_{i}^*(\br_\mu)\varphi_k(\br_\mu)\left[L_{\mu\gamma} - L_{\mu\gamma}^*\right] &=
    2 i \varphi_{i}^*(\br_\mu)\varphi_k(\br_\mu)\mathrm{Im}\left[L_{\mu\gamma}\right]. 
    \end{align}
The storage requirement for any $\hat{v}_{\gamma\pm}$ in again only $\mathcal{O}(M^2)$.
We see that an added advantage of this ISDF form is that all operations involved in propagating a walker can be now be performed using the real matrices $\mathrm{Re}\left[L_{\mu\gamma}\right]$ and $\mathrm{Im}\left[L_{\mu\gamma}\right]$ saving a factor of two in speed.

Likewise, the force-bias potential can now be evaluated as

\begin{align}
    \bar{x}_{\gamma+} &= \sum_{ik\sigma\mu}\varphi_{i}^*(\br_\mu)\varphi_k(\br_\mu)\mathrm{Re}\left[L_{\mu\gamma}\right]G_{i\sigma k\sigma}\\
                      &= \sum_{\mu\sigma} \mathrm{Re}\left[L_{\mu\gamma}\right]\tilde{G}_{\mu\sigma\mu\sigma},
\end{align}
where
\begin{equation}
    \tilde{G}_{\mu\sigma\nu\sigma} = \sum_{ik}\varphi_{i}^*(\br_\mu)G_{i\sigma k\sigma}\varphi_k(\br_\nu),
\end{equation}
and similarly
\begin{equation}
\bar{x}_{\gamma-} = \sum_{\mu\sigma} \mathrm{Im}\left[L_{\mu\gamma}\right]\tilde{G}_{\mu\sigma\mu\sigma}.
\end{equation}
The above steps can be efficiently calculated using dense matrix-matrix multiplications and the storage requirement for any intermediate matrices never exceeds $c_\mu^2 M^2$.

Some care needs to be taken when evaluating the local energy via \cref{eq:local_energy}.
There are two concerns: 1) the order of contractions when constructing intermediate tensors; 2) the size of the prefactor associated with these operations which depends on $c_\mu$.
We will first focus on reducing the prefactor.

Consider the expression for the `half-transformed' ERI tensor
\begin{equation}
	\tilde{v}_{abkl}^{\sigma\sigma'} = \sum_{ij} v_{ijkl} [V^*_\sigma]_{ia} [V^*_{\sigma'}]_{jb},
\end{equation}
which appears in the evaluation of the local energy.
We could proceed and compute $v_{abkl}$ using the existing ISDF orbital products constructed using the full set of $M$ orbitals.
However, it has been found previously that total energies typically converge faster with respect to $c_\mu$ when ISDF is performed on orbital products containing only the occupied set of orbitals, or at least with one set of occupied orbitals and one set of virtual orbitals\citep{hu2018accelerating}.
Thus, we instead perform a second ISDF factorization on the `half-transformed' orbital product set

\begin{align}
    \tilde{\varphi}^*_a(\br) \varphi_k(\br)&= \left(\sum_i \varphi_i^* (\br)[V_T^*]_{ia}\right)\varphi_k(\br)\\
                    &\approx \sum_{\mu}^{\tilde{N}_\mu} \tilde{\zeta}_\mu(\br)\tilde{\varphi}^*_a(\br_\mu) \varphi_k(\br_\mu)\label{eq:half_trans},
\end{align}
where $\tilde{N}_\mu = c_E M$.
To distinguish the two approaches, in what follows we will denote by $c_P$ as the rank parameter of the ISDF procedure for the full set of orbitals which is used for propagating the walkers and $c_E$ as the rank parameter used for the half-transformed set, which are used to compute the walker's energy.
Inserting \cref{eq:half_trans} into \cref{eq:four_ix} we can identify

\begin{align}
    \tilde{M}_{\mu\nu} = \frac{4\pi}{\Omega} \sum_{\bG\ne\mathbf{0}} \frac{\tilde{\zeta}_\mu(\bG)\tilde{\zeta}_\nu(-\bG)}{\bG^2}.\label{eq:muv_sym}
\end{align}
Note that due to the half-transformation, $\tilde{M}_{\mu\nu}$ is symmetric but not Hermitian.

Writing $E_{2B}=E_\mathrm{C} + E_{\mathrm{X}}$ we can straightforwardly compute
\begin{equation}
    E_\mathrm{C} = \frac{1}{2}\sum_{\sigma\sigma'}\sum_{\mu\nu} \tilde{\mathcal{G}}^{\sigma\sigma}_{\mu\mu}\tilde{M}_{\mu\nu}\tilde{\mathcal{G}}^{\sigma'\sigma'}_{\nu\nu},
\end{equation}
where we have introduced
\begin{equation}
    \tilde{\mathcal{G}}_{\mu\sigma\nu\sigma} = \sum_{ak}\tilde{\varphi}_{a}^*(\br_\mu)\mathcal{G}_{a\sigma k\sigma}\varphi_k(\br_\nu),
\end{equation}
Similarly the exchange energy can be calculated as
\begin{align}
    E_{\mathrm{X}} &= -\frac{1}{2}\sum_{\sigma}\sum_{\mu\nu} \tilde{\mathcal{G}}^{\sigma\sigma}_{\mu\nu} \tilde{M}_{\mu\nu} \tilde{\mathcal{G}}^{\sigma\sigma}_{\nu\mu} \\
    &=-\frac{1}{2}\sum_{\sigma}\sum_{\mu\nu} T^{\sigma\sigma}_{\mu\nu} \tilde{\mathcal{G}}^{\sigma\sigma}_{\nu\mu}
    \label{eq:thc_exx}.
\end{align}
The most expensive step comes from the $\mathcal{O}(c_E^2 N M^2)$ cost of constructing $\tilde{\mathcal{G}}_{\mu\sigma\nu\sigma}$.

\begin{figure}
    \includegraphics[width=0.75\textwidth]{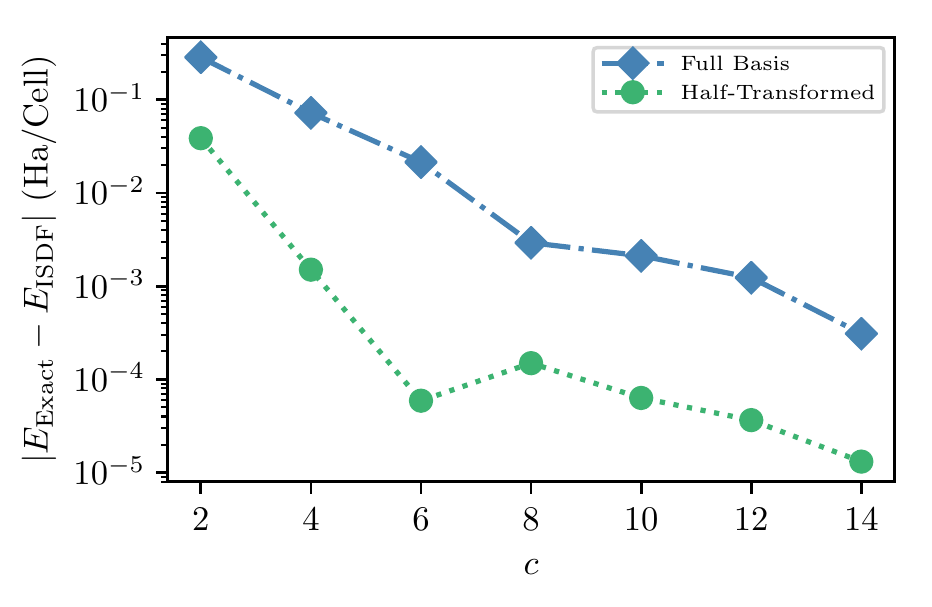}
    \caption{Comparison between the convergence of the Hartree--Fock energy when using the full set of orbital products (blue diamonds) to the half-transformed set (green circles) for the ISDF procedure. All energies were computed for a $2\times2\times2$ supercell of diamond in the DZVP basis set with $a=3.6$ \AA. Lines joining the points are meant as guides to the eye. \label{fig:hartree_fock}}
\end{figure}

\begin{figure}
    \includegraphics[width=0.75\textwidth]{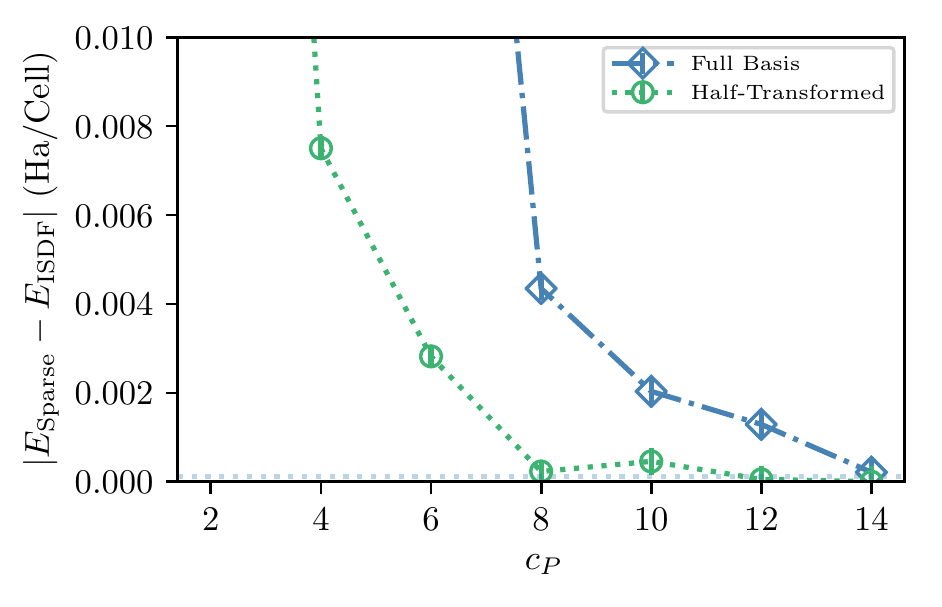}
    \caption{Comparison between the convergence of the AFQMC total energies calculated using ISDF with the full set of orbital products (blue diamonds) and the half-transformed set (green circles). The energies are plotted as a function of the ISDF rank parameter used to construct the propagator, $c_P$, whilst we fixed $c_E$ at $10$ for the half-transformed case. $E_{\mathrm{Sparse}}$ denotes the AFQMC total energy calculated using the traditional modified Cholesky decomposition for factorizing the ERIs. The system shown here is a $2\times2\times2$ supercell of diamond (16 atoms) using the DZVP basis set corresponding to 64 electrons in 208 MOs. All energies were computed for $a=3.6$ \AA. Lines joining the points are meant as guides to the eye. 1-$\sigma$ error bars in the sparse AFQMC energy are given by the horizontal dotted line.\label{fig:half_transform}}
\end{figure}

In \cref{fig:hartree_fock} we demonstrate the benefit of performing a second ISDF factorization for the computation of the energy. We compare the convergence of the Hartree--Fock energy of a $2\times2\times2$ supercell of diamond computed using ISDF procedure performed on the full MO basis compared the half-transformed one. We see that in the half-transformed case the HF energy converges much faster than the full MO case.
We find a similar trend in the corresponding AFQMC total energies as seen from \cref{fig:half_transform}. Typically we find a value of $c_E \approx 8$ and $c_P\approx 15$ is sufficient to converge the AFQMC energy to within less that one mHa per atom, corresponding to a factor of four saving in the evaluation of the total energy.

\begin{figure}
    \includegraphics[width=0.75\textwidth]{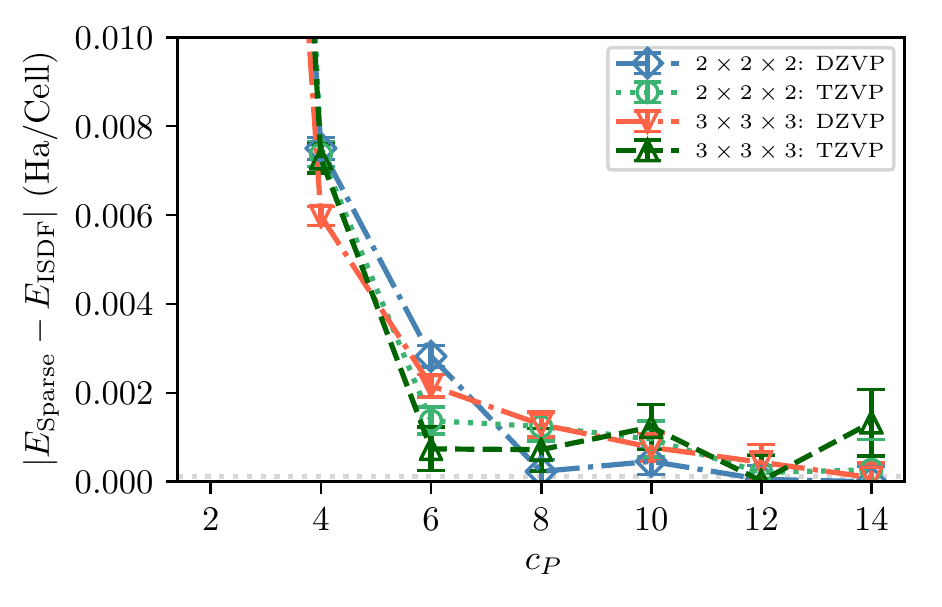}
    \caption{Convergence of the AFQMC total energy as a function of the ISDF rank parameter used to construct the propagator, $c_P$, for different supercell and basis set sizes. The ISDF rank parameter for the energy evaluation $c_E$ was fixed at 10 for all simulations. The energy differences are measured with respect to the AFQMC total energy using the traditional sparse approach. All energies were computed for $a=3.6$ \AA. Lines joining the points are meant as guides to the eye. 1-$\sigma$ error bars in the sparse AFQMC energy are given by the horizontal dotted line.\label{fig:size_conv}}
\end{figure}

We next investigate the size dependence of the ISDF rank parameters. As in previous applications of ISDF, we find that they are roughly independent of system size.
This is shown in \cref{fig:size_conv} where we compare the convergence of the AFQMC total energy of diamond for different supercell and basis set sizes.


\subsection{Algorithmic Scaling Summary}

In \cref{tab:scale} we summarize the dominant scaling of the AFQMC algorithm in terms of computational cost and memory overhead using the different approaches outlined in the previous sections.
As a rule of thumb we suggest that the sparse approach is best for simulations of small unit cells with a large $k$-point grid, due to smaller prefactor in the energy evaluation.
However, the ISDF approach is favourable nearly everywhere else and as an added benefit does not require the computationally expensive setup cost associated with constructing the tensor in \cref{eq:tensor_v} which can be significant in large scale simulations.

\begin{figure}
    \includegraphics[width=1\textwidth]{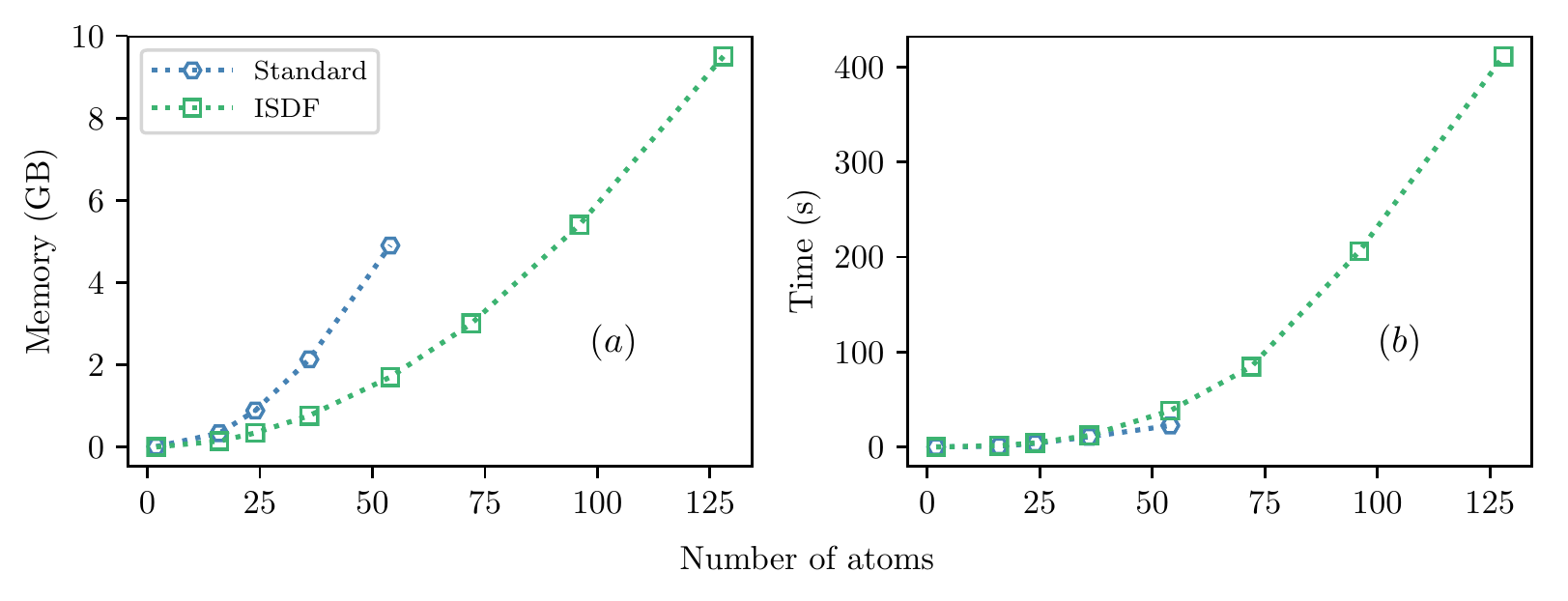}
    \caption{Comparison between the sparse modified Cholesky (labelled `Standard' above) and the ISDF approaches for the (a) memory overhead and (b) computational cost as a function of the number of atoms for diamond in the DZVP basis. Each carbon atom contributes four electrons and 13 basis functions. For the ISDF calculation we fixed $c_P=15$ and $c_E=10$ and for the standard calculation we used a Cholesky convergence criteria of $1\times 10 ^{-5}$. Running times were estimated from the time to propagate 12 walkers ten imaginary time steps with the workload distributed among 36 cores (Intel Xeon E5-2695). \label{fig:timing}}
\end{figure}

To give an idea of the actual cost of the simulations presented here, in \cref{fig:timing} we compare the memory usage and the computational cost in the DZVP basis set. We note that in practice the standard approach is only moderately faster for smaller system sizes, and is impractical for more than 54 carbon atoms in this basis set. 
\begin{table}
    \centering
    \begin{tabular}{lcc}
        \hline
        Approach & Memory & Computation \\
        \hline
        Standard & $\mathcal{O}(N^2 M^2)$ or $\mathcal{O}(M^3)$ & $\mathcal{O}(N^2 M^2)$\\
        Sparse & $\mathcal{O}(N M^2)$ & $\mathcal{O}(N M^2)$\\
        ISDF & $\mathcal{O}(M^2)$ & $\mathcal{O}(N M^2)$\\
        \hline
    \end{tabular}
    \caption{Comparison between the fundamental memory and computational scaling of the AFQMC algorithm using the standard, sparse (using $k$-point symmetry) and the ISDF approaches.\label{tab:scale}}
\end{table}
\FloatBarrier

\section{Results}
In this section we apply these developments to compute the structural properties of diamond.
All simulations were performed using Goedecker-Teter-Hutter
(GTH)\citep{GTH1996} type pseudo-potentials constructed with the
Perdew-Burke-Ernzerhof (PBE)\citep{PBE1996} exchange-correlation functional and its associated Gaussian basis sets, as supplied by the CP2K\citep{doi:10.1063/1.2770708,CP2K2013} software package.
Periodic Hartree--Fock calculations were performed with the PySCF package\citep{PYSCF} in order to generate the MO coefficients and  periodized atomic orbitals necessary for the ISDF factorization. 
All AFQMC calculations were performed using the open-source QMCPACK software package\citep{qmcpack}.
The ISDF factorization was computed with an in-house code which will in future be distributed with QMCPACK.
We used $\sim$200 walkers and timesteps of 0.0025 Ha$^{-1}$ for convergence calculations and 0.005 Ha$^{-1}$ for the calculation of the equation of state of Carbon (see Supporting Information).

\begin{figure}
    \includegraphics[width=0.75\textwidth]{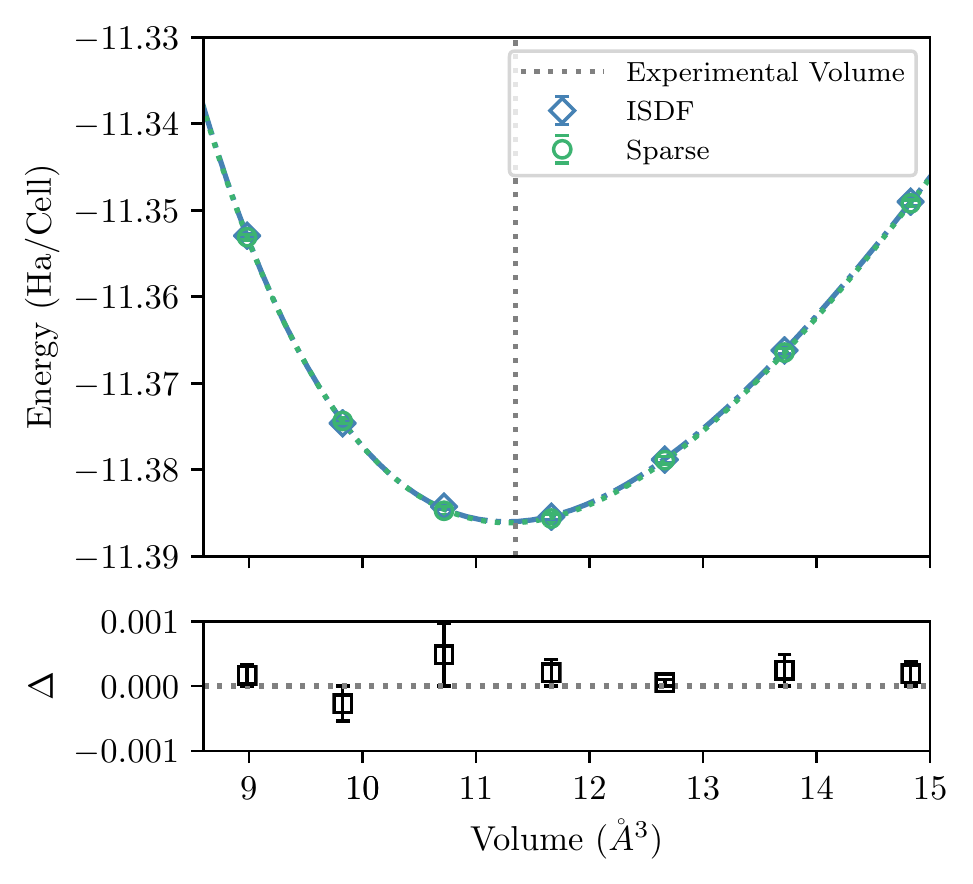}
    \caption{Top Panel: Comparison between the AFQMC equation of state calculated with the traditional approach using the modified Cholesky decomposition (denoted here as Sparse) and the ISDF approach for a $3\times3\times3$ supercell of diamond in the TZVP basis set. Lines joining the points are Vinet equation of state fits to the data \citep{0953-8984-1-11-002,jones2014scipy}. The AFQMC error bars, where not visible, are smaller than the symbols. Bottom Panel: Error in the ISDF energies (\mbox{$\Delta = E_{\mathrm{Sparse}}-E_{\mathrm{ISDF}}$ (Ha/Cell))} as a function of the unit cell volume. \label{fig:cold_curve}}
\end{figure}

In \cref{fig:cold_curve} we present our AFQMC calculations for the equation of state of a $3\times3\times3$ supercell of diamond in the TZVP basis set.
We find perfect agreement between the existing sparse-linear algebra approach the ISDF approach introduced here, with all points lying within error bars of each other.

\cref{tab:eqlb} summarizes our results for the equilibrium lattice constant ($a_0$) and bulk modulus ($B_0$) computed from a Vinet\citep{0953-8984-1-11-002} equation of state fit to the data in \cref{fig:cold_curve}.
We compare our results to those computed using other quantum chemical methods (HF, MP2 and CCSD) using the same single-particle basis set and pseudo-potential.
Overall we find that the AFQMC results compare well with the experimental values and CCSD results.

Also presented in \cref{tab:eqlb} is the AFQMC result for the cohesive energy of diamond ($\Delta E = E_{\mathrm{atom}} - E_{\mathrm{solid}}$) computed at the $T=300$ K experimental lattice constant ($a=3.567$\AA) in the TZVP basis set.
We computed the AFQMC cohesive energy by first extrapolating the AFQMC correlation energy from the $3\times3\times3$ and $4\times4\times4$ supercells to the infinite system size limit assuming a $N_k^{-1/3}$ dependence of the correlation energy\citep{mcclain_gaussian-based_2017}.
The $4\times4\times4$ supercell corresponds to a direct simulation of 256 electrons in 2176 orbitals and represents an impractically large calculation using the regular sparse approach. In contrast, the ISDF factorization required just 17 GB of memory to store the ERIs and was run on 1152 cores for roughly 16 hours to get the error bars quoted here of $<1$ mHa/Cell.
We used the separately converged (with respect to $N_k$) Hartree--Fock energy from Ref.\citenum{mcclain_gaussian-based_2017} of -11.0965 Ha/Cell to give an estimate of -11.407(7) Ha/Cell for the AFQMC total energy.
The error bar is systematic in nature and is estimated from the system size extrapolation (see Supplementary Information for further details.)

For the atomic calculation we constructed the trial wavefunction from the UHF solution for the triplet ground state of a carbon atom with the same pseudo potential used in the solid state calculation.
We performed a basis-set extrapolation of the AFQMC atomic energy using the GTH-CC-TZVP and GTH-CC-QZVP basis sets to find \mbox{$E_{\mathrm{atom}}=-5.448(6)$ Ha}, where the error bar was estimated from the basis set extrapolation.
The atomic energies found by using the same DZVP and TZVP basis sets used in the solid state calculations gave similar results.
We find that the AFQMC result for the cohesive energy agrees well with the CCSD value and experiment.
Residual finite-size and basis set errors probably explain the remaining discrepancy to experiment.
For example, the AFQMC cohesive energy changes by +0.25 eV/atom when moving from the DZVP to TZVP basis sets.
A similar increase when going from TZVP to the complete basis set limit would bring the AFQMC energy in even better agreement with experiment.
We should stress that in addition to the basis set extrapolation, a more careful analysis would involve twist averaging and the application of finite size corrections\citep{LinTA01,FraserFS96,ChiesaRPA06,KZK2008,Drummond2008,KZK2011,HolzmannFS16}. Further information on the cohesive energy calculation is availible in the Supporting Information.

\begin{table}
    \centering
    \begin{tabular}{lccc}
        \hline
        & $a_0$ (\AA) & $B_0$ (GPA) & $\Delta E$ (eV/atom) \\
        \hline
        HF & 3.527 & 507 & 5.36 \\
        MP2 & 3.545 & 436 & 7.91 \\
        CCSD & 3.539 & 463 & 7.04 \\
        AFQMC (Sparse) & \hspace{1em} 3.561(2) & 441 & - \\
        AFQMC (ISDF) & \hspace{1em} 3.559(2) & 442 & \hspace{1em} 6.95(19) \\
        Experiment & 3.553 & 455 & 7.55 \\
        \hline
    \end{tabular}
    \caption{AFQMC results for the equilibrium lattice constant and bulk modulus ($B_0$) of diamond compared to the HF, MP2 and CCSD results from Ref.\citenum{mcclain_gaussian-based_2017} using a $3\times3\times3$ $\Gamma$-centered Monkhorst-Pack grid\citep{monkhorst_pack}. All methods used the same pseudo-potential and TZVP basis set. The AFQMC results for the bulk modulus and equilibrium lattice constant were determined from a Vinet fit to the data\citep{0953-8984-1-11-002,jones2014scipy}. 1-$\sigma$ error bars from the fit are in the last digit in $a_0$ and are given by the number in parenthesis; they are not significant in the case of $B_0$. Experimental values have been corrected fo zero-point vibrational effects\citep{KresseHSESol2011}. Cohesive energies were computed in the infinite system size limit but using the same TZVP basis set. The error bar in the AFQMC cohesive energy give an estimate of the remaining systematic basis set and system size errors.\label{tab:eqlb}}
\end{table}
\section{Conclusions}
In conclusion, we have introduced the use of ISDF as a means to reduce the memory bottleneck associated with storing the ERIs in AFQMC simulations of real materials in Gaussian basis sets.
Using this approximation allows us to reduce the memory storage requirement of AFQMC from $\mathcal{O}(M^4)$ to $\mathcal{O}(M^2)$, significantly extending the scope of the algorithm.
In particular, direct simulations of solid state systems containing more that 2000 basis functions are now possible using this approach.
With this development, we provided benchmark AFQMC results for the structural properties of diamond allowing for a direct comparison to other quantum chemical methods.
A significant advantage of the ISDF approach is that it allows the entire algorithm to use dense linear algebra and is thus well suited to a GPU implementation\citep{shee_gpu}.
Although we focused on a relatively simple solid state application we expect ISDF to be useful for other materials.
A more systematic study of a wider variety of materials is left for future investigation.

\FloatBarrier
\section{Acknowledgements}
We would like to thank Mario Motta and Joonho Lee for interesting discussions and Qiming Sun and Timothy Berkelbach for assistance in running PySCF.
FDM would like thank the Institute for Nuclear Theory at the University of Washington for its hospitality.
This work was performed under the auspices of the U.S. Department of Energy
(DOE) by LLNL under Contract No. DE-AC52-07NA27344.  Funding support was from
the U.S. DOE, Office of Science, Basic Energy Sciences, Materials Sciences and
Engineering Division, as part of the Computational Materials Sciences Program
and Center for Predictive Simulation of Functional Materials (CPSFM).  Computer
time was provided by the
Argonne Leadership Computing and Livermore Computing Facilities.
\section{Supporting Information}
More details on the basis set and system size extrapolations and population control analysis is included in the Supporting Information. This information is available free of charge via the Internet at http://pubs.acs.org.

\bibliography{refs}

\begin{mcitethebibliography}{74}
\bibitem[hyb()]{hybrid}
Note that in order to reduce the frequency of energy evaluations we use
  exclusively use the hybrid method for computing the importance function
  rather than the real local energy method\citep{zhang201315}.\relax
\mciteBstWouldAddEndPunctfalse
\mciteSetBstMidEndSepPunct{\mcitedefaultmidpunct}
{}{\mcitedefaultseppunct}\relax
\EndOfBibitem
\bibitem[fac()]{factorize_M}
Formally $M$ is a Hermitian positive definite matrix and can be Cholesky
  decomposed. In practice due to numerical issues it may become only positive
  semi-definite at which point one can use the modified Cholesky
  \citep{mchol0,mchol,mchol2,doi:10.1063/1.3654002} decomposition used in
  quantum chemistry to decompose the ERI tensor. In this work we did not
  encounter such issues and used the traditional Cholesky decomposition.\relax
\mciteBstWouldAddEndPunctfalse
\mciteSetBstMidEndSepPunct{\mcitedefaultmidpunct}
{}{\mcitedefaultseppunct}\relax
\EndOfBibitem
\end{mcitethebibliography}

\end{document}